\documentclass[reprint,pra,aps,twocolumn]{revtex4-2}
\usepackage{amsmath,amssymb,amsthm,mathtools,bm}
\usepackage{microtype}
\usepackage{enumitem}
\usepackage[colorlinks=true,linkcolor=blue,citecolor=blue,urlcolor=blue]{hyperref}
\usepackage[normalem]{ulem}

\newtheorem{theorem}{Theorem}[section]
\newtheorem{lemma}[theorem]{Lemma}
\newtheorem{proposition}[theorem]{Proposition}
\newtheorem{corollary}[theorem]{Corollary}
\theoremstyle{definition}

\newtheorem{example}[theorem]{Example}
\newtheorem{definition}[theorem]{Definition}

\newcommand{\Tr}{\operatorname{Tr}}

\newcommand{\conv}{\operatorname{conv}}
\newcommand{\rank}{\operatorname{rank}}
\newcommand{\PS}{\mathsf{PS}}
\newcommand{\PM}{\mathsf{P}}
\newcommand{\id}{\mathbb{I}}
\newcommand{\R}{\mathbb{R}}
\newcommand{\C}{\mathbb{C}}
\newcommand{\ket}[1]{\lvert #1\rangle}

\newcommand{\proj}[1]{\lvert #1\rangle\!\langle #1\rvert}
\newcommand{\norm}[1]{\left\lVert #1\right\rVert}
\newcommand{\abs}[1]{\left\lvert #1\right\rvert}

\begin{document}

\title{Robust Certification of Non-projective Qubit Measurements without Inputs}

\author{Sumit Rout}
\affiliation{International Centre for Theory of Quantum Technologies (ICTQT), University of Gda{\'n}sk, ul. prof. Marii Janion 4, 80-309 Gda{\'n}sk, Poland}
\author{Pawe{\l} Horodecki}
\affiliation{International Centre for Theory of Quantum Technologies (ICTQT), University of Gda{\'n}sk, ul. prof. Marii Janion 4, 80-309 Gda{\'n}sk, Poland}
\affiliation{Faculty of Applied Physics and Mathematics, Gda{\'n}sk University of Technology, Gabriela Narutowicza 11/12, 80-233 Gda{\'n}sk, Poland}

\author{Some Sankar Bhattacharya}
\affiliation{Física Teòrica: Informació i Fenòmens Quàntics, Universitat Autònoma de Barcelona, 08193 Bellaterra, Spain}
\begin{abstract}
Positive-operator-valued measures (POVMs) beyond projective simulation (aka non-projective measurements) offer advantages over projective measurements, motivating their certification directly from observed statistics. Here, we show that this certification can be achieved robustly without \emph{input randomness}, using only the joint outcome distribution of two measurement devices. We consider two spatially separated observers whose only shared resource is an uncharacterised two-qubit state. Each observer performs a local measurement. In the projective-simulable case, the measurement device privately randomises over qubit projective measurements and applies arbitrary classical post-processing to their outcomes. In the noiseless regime, we identify three- and four-outcome anti-correlations that certify that both local measurements lie beyond projective simulation. To establish noise robustness, we develop a non-linear {\it covariance witness} that combines the geometry of projective measurements with a contraction bound valid for every two-qubit state. For the correlations generated by identical qubit trine measurements or identical qubit SIC-POVM on noisy qubit singlet, we obtain exact critical white-noise fractions of (1/4) and (1/3), respectively. Below these thresholds, no pair of projective-simulable measurements on any two-qubit state can reproduce these correlations. We present explicit projective simulations that attain the thresholds and cover the remaining noise range. Finally, the same method yields a strictly positive noise tolerance for certification whenever both observers apply the same extremal qubit POVM with three or four non-zero effects to a depolarised qubit singlet. These results establish a general analytic framework for certifying non-projective qubit measurement resources without input randomness.
\end{abstract}

\maketitle

\section{Introduction}
In quantum mechanics, measurements are described by positive-operator-valued measures (POVMs). These are collections of positive semidefinite operators, one for each outcome, that sum to the identity~\cite{Nielsen2012}. The familiar projective measurements (PVMs), whose effects are mutually orthogonal projectors, form a special case. Classical randomisation over different projective measurements and further post-processing of their outcomes can extend the reach of PVMs on the same system, giving rise to the broader class of \emph{projective-simulable} measurements~\cite{Oszmaniec2017}. POVMs beyond this class (aka non-projective) are resourceful in quantum information processing: they offer advantages over projective simulations in state discrimination~\cite{Ivanovic87, Peres88, DiMario22, QuantumStateDisriminationRev}, entanglement detection \cite{Shang2018}, quantum tomography \cite{Derka1998, Renes2004, OptTomography, singleSettingTom, povmSHADOWGhune}, quantum metrology \cite{Ragy2016, Szcyykulska2016, rafalMETRO}, quantum cryptography \cite{optimalCRYPTO} and  port-based teleportation \cite{Ishizaka2008, Studzinski2017port, Mozrzymas2018optimal}. They also enable stronger Bell-inequality violations~\cite{Vertesi2010}, and enhance the certified randomness extraction~\cite{Acin16}. These advantages motivate the certification of measurements beyond projective simulation directly from the observed statistics.

The device-independent framework offers a natural approach to certification with minimal device knowledge. Treating the devices as black boxes, it draws conclusions from their observed correlations under specified operational assumptions~\cite{Brunner14}. For certification beyond projective simulation, however, one must also account for the dimension of the underlying system. It follows from Naimark’s theorem that one can realise any POVM by performing a projective measurement on a larger system. With additional classical randomisation over PVMs and outcome post-processing, an auxiliary system of the same dimension as the measured system suffices~\cite{Oszmaniec2017}. Distinguishing a non-projective measurement from such a projective implementation therefore requires specifying the system to which the certification applies. An upper bound on its dimension supplies this information in the semi-device-independent framework, while leaving the states and measurement operators uncharacterised.

Such semi-device-independent certification schemes have been developed  in Bell experiments, using correlations between local measurements on shared entangled states~\cite{Vertesi2010,Gomez2016,Acin16,Gomez2018}, and in prepare-and-measure experiments with dimension-bounded systems. In the latter setting, Tavakoli et al.\ developed methods for robust certification and self-testing of three- and four-outcome qubit POVMs~\cite{Tavakoli20}, while Mart\'inez et al.\ certified a seven-outcome measurement beyond projective simulation in dimension four~\cite{Martinez23}. These schemes use externally chosen preparation and/or measurement settings. The input choices must satisfy appropriate independence conditions, making input randomness part of the certification resources. Although quantum processes can provide intrinsic randomness, establishing the unpredictability of their outputs requires trusted characterisation or a separate certification procedure ~\cite{Bierhorst2018}. Certification scheme without external inputs would remove this additional randomness requirement. Renou et al. demonstrated quantum nonlocality in a triangle network with independent sources and fixed local measurements~\cite{Renou19}. Earlier works also established randomness-free tests of non-classicality using a fixed source and fixed local measurements, certifying quantum correlations beyond those achievable with classical systems of the same local dimension or, equivalently, by classically post-processing the outcomes of a  \emph{single} PVM at each site~\cite{Guha2021,Ma2023}. Subsequent work broadened these schemes for qubits and extended them to qutrits~\cite{Rout2025}. Here, we ask: can two observers sharing only
an uncharacterised two-qubit state use a single joint outcome distribution from their local measurement device to robustly certify non-projective measurement? We allow alternative implementations to use arbitrary {\it private randomisation} over local projective measurements and {\it classical post-processing} of outcomes.\\

We answer this question affirmatively. Our comparison includes the full projective-simulable class, with randomisation over PVMs and outcome post-processing allowed to depend on the selected PVM. The two-qubit state is the observers' only shared resource, the private random choices of local measurement devices  are independent of each other and of the source. This restriction is essential: with unrestricted shared classical randomness, the devices could jointly sample any prescribed pair of outputs and reproduce every joint distribution. We first derive noiseless criteria for certifying three- and four-outcome qubit non-projective measurements at both sites. We then develop an analytic framework for noise-robust certification. Our central tool is a nonlinear \emph{covariance witness} whose bound follows from the geometry of projective-simulable measurements and holds for every shared two-qubit state. A violation certifies that at least one local measurement lies beyond projective simulation. For trine and tetrahedral SIC-POVMs on a two-qubit singlet mixed with white noise, we obtain exact critical noise fractions of $1/4$ and $1/3$, respectively. Below these thresholds, no pair of projective-simulable measurements on any two-qubit state can reproduce the correlations. Explicit projective simulations attain the thresholds and reproduce the correlations throughout the remaining noise range. Finally, the same method also establishes a strictly positive depolarising-noise tolerance for correlations generated by each genuine three- or four-outcome extremal qubit POVM.\\

Section \ref{sec:preliminaries} introduces projective-simulable measurements, reviews relevant properties of qubit POVMs and specifies our certification scenario. We then establish the noiseless certification criteria for three- and four-outcome measurements in Section \ref{sec:noiseless}. In Section \ref{sec:noisy}, we develop the covariance witness and apply it to obtain exact noise thresholds for qubit trine and SIC POVM, together with robustness bounds for genuine three- and four-outcome extremal qubit POVMs. We conclude with a discussion of implications and open questions in Section \ref{sec:discussion}. The technical proofs appear in the appendices.

\section{Preliminaries}
\label{sec:preliminaries}

Let $\mathcal H\simeq\C^d$ be the Hilbert space of a $d$-dimensional quantum system. An $n$-outcome POVM is a collection $M=\{M_a\}_{a\in\Omega_n}$
of effects satisfying \(M_a\ge0\) and \( \sum_a M_a=\id_d\), where $\Omega_n=\{0,\ldots,n-1\}$ is the
outcome set and $\id_d$ is the identity operator on $\C^d$.
For a state $\rho$, the probability of outcome $a$ is $\Tr(\rho M_a)$ when performing the measurement $M$. We denote the convex set of all $n$-outcome POVMs on a $d$-dimensional system by
$\mathcal P(d,n)$. Stochastically mixing measurements $M$ and $N$ yields a POVM with effects: $[tM+(1-t)N]_a=tM_a+(1-t)N_a$, where $0\le t\le1$.
A projective measurement (PVM) has mutually orthogonal projector effects \(\{\Pi_a\}_{a\in\Omega_n}\),
$\Pi_a\Pi_b=\delta_{ab}\Pi_a~ \forall a,b\in\Omega_n$.
We write $\PM(d,n)$ for the set of PVMs on a $d$-dimensional system with outcomes in $\Omega_n$,
allowing zero effects so that measurements with fewer outcomes can be represented on the same outcome set.\\

\begin{figure*}[t]
    \centering
    \includegraphics[scale=0.198]{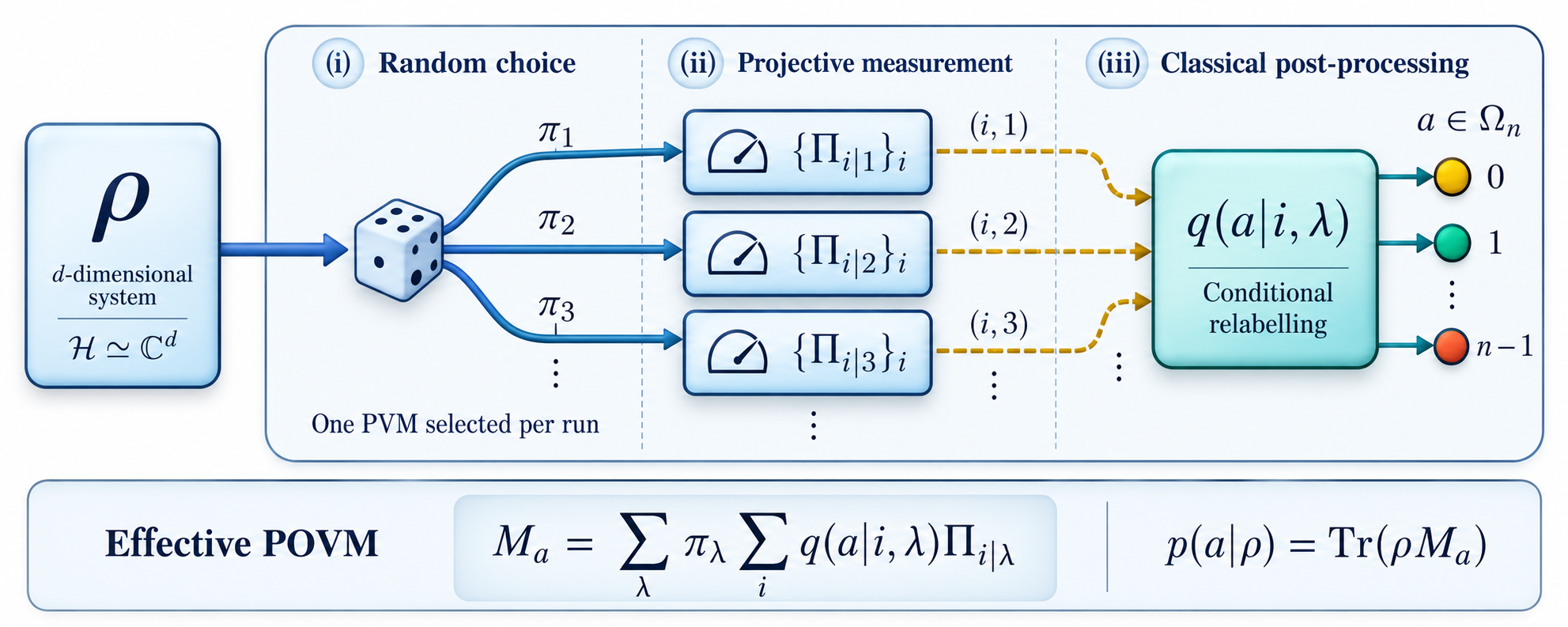}
    \caption{A schematic representation of a projective-simulable measurement on a $d$-dimensional system.  First, a PVM $\{\Pi_{i|\lambda}\}_i$ is selected with probability $\pi_\lambda$, and its outcome $i$ is then mapped to a final label $a$ with probability $q(a|i,\lambda)$. The classical post-processing of the observed outcome may depend on the chosen PVM. The resulting POVM has effects $M_a=\sum_\lambda\pi_\lambda\sum_i q(a|i,\lambda)\Pi_{i|\lambda}$.}
    \label{fig:projective-simulation}
\end{figure*}

Classical  randomisation and outcome post-processing expand the class of measurements implementable with PVMs on the same $d$-dimensional system.
As illustrated in Fig.~\ref{fig:projective-simulation}, a projective simulation may randomly choose a PVM, measure it, and assign a final label to the observed outcome according to a conditional probability distribution. The assignment may depend on the chosen PVM.

\begin{definition}[Projective-simulable measurement~\cite{Oszmaniec2017}]
\label{def:PS}
A POVM $M\in\mathcal P(d,n)$ is \emph{projective-simulable} if
\begin{equation}
  M_a=\sum_\lambda \pi_\lambda
      \sum_i q(a|i,\lambda)\Pi_{i|\lambda},
  \label{eq:prelim-ps-model}
\end{equation}
where each $\{\Pi_{i|\lambda}\}_i$ is a PVM on $\C^d$,
$\pi_\lambda\ge0$, $\sum_\lambda\pi_\lambda=1$, and
$q(a|i,\lambda)\ge0$, $\sum_a q(a|i,\lambda)=1$ for all $i,\lambda$.
We denote this class by $\PS(d,n)$.
\end{definition}

In our earlier work on qubit and qutrit measurements~\cite{Rout2025}, we considered simulations in which each device classically post-processes the outcomes of a single fixed PVM. This corresponds to each device assigning probability one to a single PVM in Eq.~\eqref{eq:prelim-ps-model}. Here we certify qubit measurements against the full class $\PS(2,n)$. We test whether the observed correlations can be reproduced by a two-qubit state and two local POVMs in $\PS(2,n)$. Each device may privately randomise over qubit PVMs and choose its post-processing rule according to the selected PVM.

\begin{lemma}[Convex-hull characterisation~\cite{Oszmaniec2017}]
\label{lem:convhull}
For finite $d$ and $n$, \(\PS(d,n)=\conv\PM(d,n)\), where \(\conv\PM(d,n)\) is the convex hull of measurements in \(\PM(d,n)\).
\end{lemma}

The set $\PM(d,n)$ is compact, so its convex hull is compact in the
finite-dimensional space of Hermitian operators. Consequently, allowing continuous
randomisation does not enlarge $\PS(d,n)$. The measurements on a $d$-dimensional system which are not projective simulable are referred to as \emph{non-projective} measurements.
A POVM $M\in \mathcal P(d,n)$ is \emph{extremal} if it cannot be expressed as a nontrivial
convex mixture of distinct POVMs in $\mathcal P(d,n)$. The preceding
characterisation yields the following corollary, which we use in our noiseless proofs.
\begin{corollary}
\label{cor:extremalPS}
An extremal POVM is projective-simulable if and only if it is a PVM.
\end{corollary}

Indeed, an extremal POVM cannot be a nontrivial convex mixture of PVMs.\\

\textit{Qubit POVMs and extremality}: For qubits, the effects of a POVM $M=\{M_a\}_{a\in\Omega_n}$ can be written as
\begin{equation}
  M_a=\alpha_a\bigl(\id_2+\bm n_a\cdot\bm\sigma\bigr),
  \qquad \alpha_a=\frac{\Tr M_a}{2}.
  \label{eq:prelim-qubit-bloch}
\end{equation}
where $\bm\sigma=(\sigma_x,\sigma_y,\sigma_z)$ is  the Pauli vector and $\bm n_a=(n_{a,x},n_{a,y},n_{a,z})$ is a real Bloch vector. For a zero effect, we set $\bm n_a=0$. Positivity and completeness are then equivalent to the following conditions
\begin{equation}
  \begin{gathered}
    \alpha_a\ge0,\qquad \norm{\bm n_a}_2\le1,\\
    \sum_a\alpha_a=1,\qquad \sum_a\alpha_a\bm n_a=0.
  \end{gathered}
  \label{eq:prelim-qubit-normalisation}
\end{equation}
For a non-zero effect, rank one is
equivalent to $\norm{\bm n_a}_2=1$.
A non-trivial qubit PVM  has exactly two non-zero effects
\begin{equation}
  \Pi_\pm(\bm m)=\frac12
       \bigl(\id_2\pm\bm m\cdot\bm\sigma\bigr),
  \qquad \norm{\bm m}_2=1  \label{eq:prelim-qubit-pvm}
\end{equation}

while a trivial qubit PVM has a single non-zero effect $\id_2$. Every binary qubit POVM can be realised by post-processing the outcomes of a PVM that diagonalises
one of its effects. Moreover, qubit PVMs  with at most two non-zero effects can simulate all measurements in $\PS(2,n)$ through classical randomisation~\cite{Oszmaniec2017}.\\

A rank-one POVM is extremal exactly when its non-zero effects are linearly
independent~\cite{DAriano2005}. Apart from the deterministic measurements, the
extremal qubit POVMs have at most four non-zero effects, all of rank one.
Those with two non-zero effects are projective. A qubit POVM with three non-zero rank-one effects is extremal if and only if its effects are pairwise non-proportional. A four-outcome rank-one qubit POVM is extremal if and only if its effects are linearly independent. In Bloch form, the latter condition is equivalent to linear independence of the four real vectors $(1,\bm n_a^T)^T$. The Bloch vectors of genuine three- and four-outcome extremal
qubit POVMs respectively span a plane and $\R^3$~\cite{DAriano2005,Tavakoli20}.
By Corollary~\ref{cor:extremalPS}, both classes lie outside $\PS(2,n)$.\\

Two symmetric examples of such qubit POVMs will be used throughout. The \emph{trine} POVM has
three effects with parameters $\alpha_a=1/3$ and
$\bm n_a\cdot\bm n_b=-1/2$ for $a\ne b$.
The \emph{tetrahedral} symmetric informationally complete (SIC) POVM
has four effects with
$\alpha_a=1/4$ and $\bm n_a\cdot\bm n_b=-1/3$ for $a\ne b$.
In both cases the Bloch vectors are unit vectors. We sometimes express planar Bloch vectors using two coordinates in an orthonormal basis of their plane, leaving the zero perpendicular component implicit.\\

\textit{Certification scenario}: Spatially separated non-communicating parties Alice and Bob share an arbitrary uncharacterised state $\rho_{AB}\in \mathcal D(\C^2\otimes\C^2)$ and perform fixed local POVMs
$M_A=\{A_a\}_{a\in\Omega_n}$ and $M_B=\{B_b\}_{b\in\Omega_n}$, respectively, with joint outcome probabilities
\begin{equation}
  p(a,b)=\Tr\!\left[\rho_{AB}(A_a\otimes B_b)\right].
  \label{eq:prelim-joint-probability}
\end{equation}
Their only shared resource is the two-qubit state. Each  measurement device may use private randomness and arbitrary local classical post-processing. The devices’ private random variables are independent of one another and jointly independent of the source. There are no external preparation or additional inputs.\\

\begin{figure*}[t]
    \centering
    \includegraphics[scale=0.2]{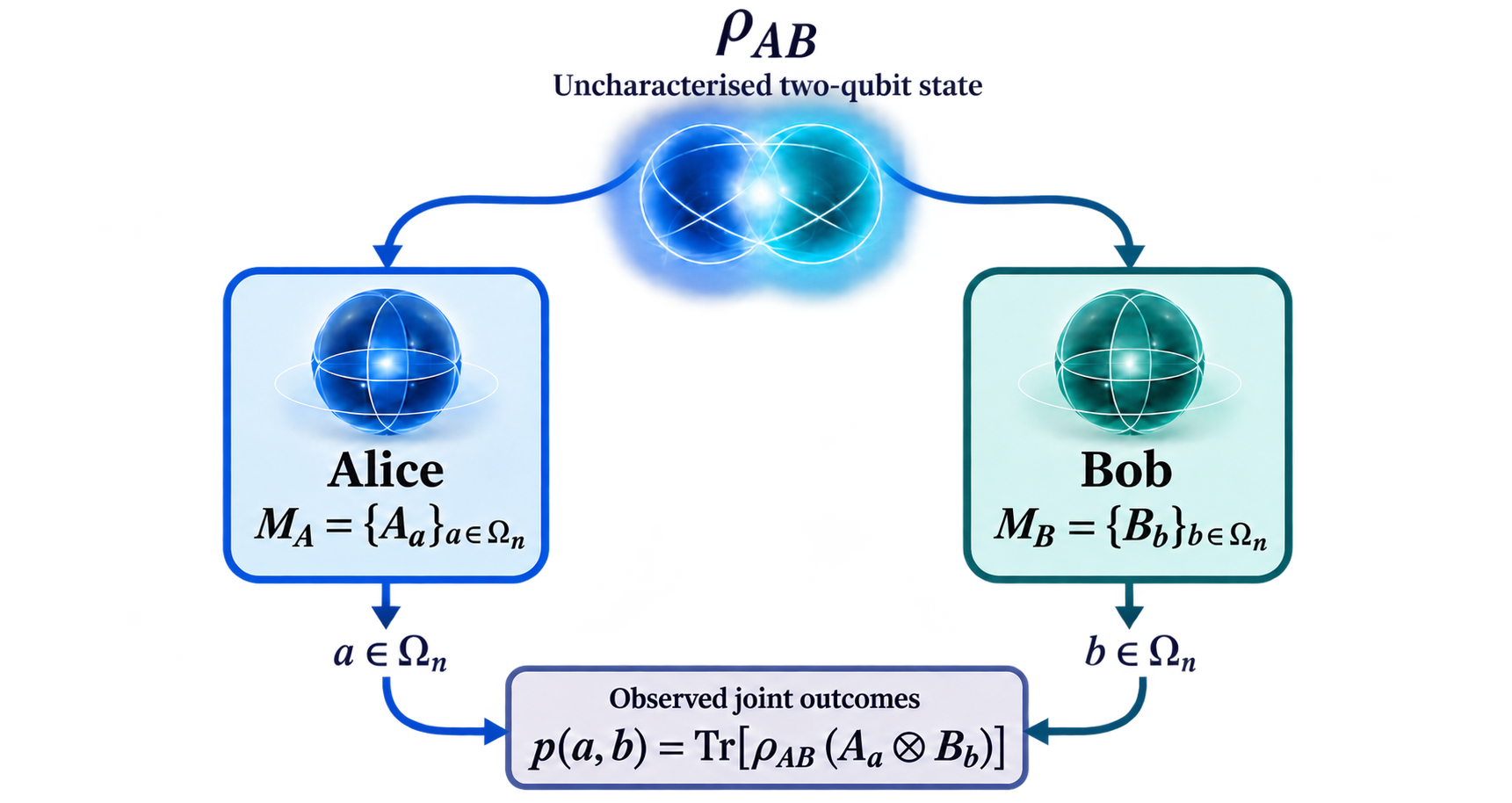}
    \caption{{\it Certification scenario:} Spatially separated non-communicating parties Alice and Bob share an uncharacterised two-qubit state $\rho_{AB}$ as their only shared resource and perform fixed local POVMs
    $M_A=\{A_a\}_a$ and $M_B=\{B_b\}_b$, {\it without external inputs}. Their outcomes follow the joint distribution $p(a,b)=\operatorname{Tr}[\rho_{AB}(A_a\otimes B_b)]$. The scheme certifies that at least one local measurement is not projective-simulable by excluding every realisation with two
    projective-simulable local measurements on any two-qubit state,
    even when arbitrary private randomisation over local qubit PVMs and local
    classical post-processing are allowed. The devices’ private random variables are independent of one another and jointly independent of the source.}
    \label{fig:setup}
\end{figure*}

We write $P=[p(a,b)]$ for the joint probability matrix and 
\begin{equation}
  p_A(a)=\sum_b p(a,b),\qquad
  p_B(b)=\sum_a p(a,b)
  \label{eq:prelim-marginals}
\end{equation}
for their respective marginals. A \emph{projective-simulable realisation} is any representation of $P$ through Eq.~\eqref{eq:prelim-joint-probability} with
$M_A,M_B\in\PS(2,n)$ and an arbitrary two-qubit state.
Excluding every such realisation certifies that at least one local
measurement is not projective-simulable.  Our noiseless criteria additionally establish the stronger conclusion that both measurements are non-projective.\\

A two-qubit state has the Bloch representation 
\begin{equation}
  \begin{aligned}
  \rho_{AB}=\frac14\Bigl[&\id_2\otimes\id_2
      +(\bm r\cdot\bm\sigma)\otimes\id_2
      +\id_2\otimes(\bm s\cdot\bm\sigma)\\
      &+\sum_{j,k=1}^3 T_{jk}\sigma_j\otimes\sigma_k\Bigr],
  \end{aligned}
  \label{eq:prelim-state-bloch}
\end{equation}
where $r_j=\Tr[\rho_{AB}(\sigma_j\otimes\id_2)]$ and
$s_k=\Tr[\rho_{AB}(\id_2\otimes\sigma_k)]$ are the local Bloch
components, and $T_{jk}=\Tr[\rho_{AB}(\sigma_j\otimes\sigma_k)]$.
Here, $\bm r,\bm s$ are the Bloch vectors of the reduced states, while $T$  is the real $3\times3$ matrix of correlations between the local Pauli observables. Upon subtracting the products of the local means gives the centred Pauli correlation matrix $K=T-\bm r\bm s^T$, with entries $K_{jk}=T_{jk}-r_js_k$.\\

To quantify robustness to noise, we use the depolarising map on qubit operators:
\begin{equation}
  \Phi_t(X)=tX+(1-t)\frac{\Tr X}{2}\id_2,
  \qquad 0\le t\le1.
  \label{eq:prelim-depolarising-map}
\end{equation}
When applied to a qubit POVM effect, this gives
\begin{equation}
  M_a^{(t)}:=\Phi_t(M_a)
    =\alpha_a\bigl(\id_2+t\bm n_a\cdot\bm\sigma\bigr),
  \label{eq:prelim-noisy-effect}
\end{equation}
where $t$ is the measurement visibility. Following ~\cite{Oszmaniec2017}, we define the critical measurement visibility by
\begin{equation}
  t_{\rm PM}(M):=
  \max\bigl\{t\in[0,1]:\Phi_t(M)\in\PS(2,n)\bigr\}.
  \label{eq:prelim-measurement-threshold}
\end{equation}
Here $\Phi_t(M)=\{\Phi_t(M_a)\}_a$. For $t=0$, the effects become $\alpha_a\id_2$. This POVM is projective-simulable, since it can be implemented by sampling $a$ with probability $\alpha_a$ independent of the state. Moreover, the convexity of $\PS(2,n)$ ensures projective simulability of $\Phi_t(M)$ for every $0\le t\le t_{\rm PM}(M)$.\\

To generate the target correlations for our certification scheme, we use the noisy two-qubit singlet as the shared state
\begin{equation}
  \begin{gathered}
    \rho_\eta=(1-\eta)\proj{\psi^-}+\eta\frac{\id_4}{4},
    \qquad 0\le\eta\le1,\\
    \ket{\psi^-}=\frac{\ket{01}-\ket{10}}{\sqrt2},
    \qquad v:=1-\eta.
  \end{gathered}
  \label{eq:prelim-noisy-singlet}
\end{equation}
Here, $\eta$ denotes the white-noise fraction and $v$ the singlet
visibility. Applying the same POVM $M$ to both sub-systems yields the joint probability distribution
\begin{equation}
  p_v^M(a,b)=\alpha_a\alpha_b
      \bigl(1-v\bm n_a\cdot\bm n_b\bigr),
  \label{eq:prelim-noisy-target}
\end{equation}
with marginals $p_A(a)=p_B(a)=\alpha_a$. The alternative realisations considered for
certification allow local qubit POVMs $M_A,M_B$ and any two-qubit state $\rho_{AB}$.\\

For any qubit state $\rho$, the depolarising map satisfies $\Tr[\Phi_t(\rho)M_a]=\Tr[\rho\Phi_t(M_a)]$. In particular, for local measurement visibilities $t_A,t_B$ on an ideal singlet,
\begin{equation}
  \Tr\!\left[\proj{\psi^-}
      \bigl(M_a^{(t_A)}\otimes M_b^{(t_B)}\bigr)\right]
  =p_{t_A t_B}^M(a,b).
  \label{eq:prelim-visibility-product}
\end{equation}
Consequently, on an ideal singlet, the target distribution $p_v^M$ is reproduced whenever $t_A t_B=v$. One may therefore use visibility $v$ on one measurement and an ideal measurement on the other, or visibility $\sqrt v$ on both measurements.

\section{Noiseless certification from anti-correlations}
\label{sec:noiseless}

We begin with certification schemes in the noiseless scenarios. We show that the absence of correlated outcomes can reveal the structure of both local
measurements. In particular, suitable three- and four-outcome
anti-correlations force each qubit measurement to be extremal and beyond
projective simulation. These criteria involve only the observed probability
matrix and apply to every two-qubit realisation of the observed distribution.

\subsection{Three-outcome non-projective measurements}
\label{sec:3outcome}

We consider three-outcome correlations in which Alice and Bob never obtain the same outcome label, while every pair of distinct labels occurs with non-zero probability:
\begin{equation}
  p(a,a)=0,\qquad p(a,b)>0\quad(a\ne b),
  \label{eq:strict-anticorr}
\end{equation}
where $a,b\in\Omega_3$. We show that every two-qubit realisation of these correlations requires both local measurements to lie beyond projective simulation.

\begin{theorem}[Three-outcome certification]
\label{thm:3out}
In every two-qubit realisation of a distribution satisfying
Eq.~\eqref{eq:strict-anticorr} both local three-outcome qubit POVMs $M_A$ and $M_B$ must be extremal. In particular, $M_A,M_B\notin\PS(2,3)$.
\end{theorem}

\begin{proof}
We first show that Alice's POVM $M_A=\{A_a\}_{a\in \Omega_3}$ has rank-one effects. The same reasoning applies to Bob's POVM $M_B=\{B_b\}_{b\in\Omega_3}$. The strictly positive off-diagonal probabilities imply that all marginal probabilities $p_A(a)>0$ and $p_B(b)>0$ for every $a,b\in\Omega_3$. Hence every local effect must be non-zero, {\it i.e.} $A_a\ne0$ and $B_b\ne0$. Suppose that some effect $A_a$ is full rank. Then $A_a\ge\mu_a\id_2$, where $\mu_a>0$ is its smallest eigenvalue. Since $B_a\ge0$, we have $A_a\otimes B_a\ge\mu_a\id_2\otimes B_a$. Taking the expectation in $\rho_{AB}$ gives 
\begin{equation}
\begin{aligned}
0=p(a,a)
&=\Tr[\rho_{AB}(A_a\otimes B_a)]\\
&\ge\mu_a\Tr[\rho_{AB}(\id_2\otimes B_a)]\\
&=\mu_a p_B(a)>0,
\end{aligned}
  \label{eq:noiseless-rank-one}
\end{equation}
which is a contradiction. Thus, every $A_a$ must have rank one. The same argument applies to Bob's effects.\\ 

No two effects within either POVM can be proportional. Indeed,
$A_a=cA_{a'}$ with $c>0$ and $a\ne a'$ would give
$0=p(a,a)=\Tr[\rho_{AB}(A_a\otimes B_{b=a})]=c\Tr[\rho_{AB}(A_{a'}\otimes B_{b=a})]=c\,p(a',a)>0$, again a contradiction. Exchanging the parties gives the same
conclusion for the effects of $M_B$. Each POVM therefore has three rank-one,
pairwise non-proportional effects and is extremal~\cite[Sec.~3, p.~5985]{DAriano2005}.
A qubit PVM has at most two non-zero effects, so
Corollary~\ref{cor:extremalPS} excludes projective simulability for both measurements.
\end{proof}

Every genuine three-outcome extremal qubit POVM $M$ generates
correlations satisfying
Eq.~\eqref{eq:strict-anticorr} when both parties apply $M_A=M_B=M$ to a two-qubit singlet state.
Its Bloch vectors are distinct unit vectors, and $\alpha_a>0$ for every $a$,  so
Eq.~\eqref{eq:prelim-noisy-target} at $v=1$ gives a probability matrix with a zero diagonal and
strictly positive off-diagonal entries. Particularly using the qubit trine POVM, this gives
the uniform anti-correlation
\begin{equation}
  p_1^{\mathrm{trine}}(a,b)=\frac{1-\delta_{ab}}{6}.
  \label{eq:noiseless-trine}
\end{equation}

\subsection{Four-outcome non-projective measurements}
\label{sec:4outcome}

The four-outcome analogue of Eq.~\eqref{eq:strict-anticorr}  does not suffice to certify non-projectivity. For four outcomes, a probability matrix with zero diagonal and strictly positive off-diagonal entries
can also arise from local mixtures of projective measurements followed by post-processing.

\begin{example}[Projective-simulable anti-correlations]
\label{ex:4counter}
Alice and Bob share an ideal two-qubit singlet. Each observer independently chooses between the $x$- and $z$-basis PVMs with equal probability. The classical post-processing assigns labels $0,1$ to the two $x$-basis outcomes
and $2,3$ to the two $z$-basis outcomes. The effective local POVM has
effects
\begin{equation}
  M_a=\frac14\bigl(\id_2+\bm n_a\cdot\bm\sigma\bigr),
  \label{eq:noiseless-mixed-pvms}
\end{equation}
where $\bm n_0=-\bm n_1=\bm e_x$ and
$\bm n_2=-\bm n_3=\bm e_z$. Applying this POVM at both sites yields the distribution
\begin{equation}
  P=\frac1{16}
  \begin{pmatrix}
    0&2&1&1\\
    2&0&1&1\\
    1&1&0&2\\
    1&1&2&0
  \end{pmatrix}.
  \label{eq:noiseless-four-counterexample}
\end{equation}
The diagonal entries vanish and all off-diagonal entries are strictly positive, although both measurements are projective-simulable. Note that the matrix has rank three.
\end{example}

The rank of $P$ supplies a further constraint for four outcomes. As we show below, full rank of $P$ requires four linearly independent effects on each side. A linear dependence among Alice's effects would produce dependent rows of $P$, while one among Bob's effects would produce dependent columns. Together with the vanishing diagonal, this gives the following criterion for certifying both local qubit measurements as non-projective.

\begin{theorem}[Four-outcome certification]
\label{thm:4out}
Let $P$ be a  $4\times4$ (four-outcome) probability matrix satisfying
\begin{equation}
  p(a,a)=0\quad(a\in\Omega_4),\qquad \rank P=4.
  \label{eq:noiseless-four-criterion}
\end{equation}
In every two-qubit realisation of distribution $P$, both local four-outcome qubit POVMs $M_A$ and $M_B$ must be extremal. In particular, $M_A,M_B\notin\PS(2,4)$.
\end{theorem}

\begin{proof}
We first show that Alice's POVM $M_A=\{A_a\}_{a\in \Omega_4}$ has rank-one effects. The same reasoning applies to Bob's POVM $M_B=\{B_b\}_{b\in\Omega_4}$. The full rank of $P$ excludes zero rows and columns. Since the entries of $P$ are non-negative, all marginal probabilities
are positive, {\it i.e.} $p_A(a)>0$ and $p_B(b)>0$ for every $a,b\in\Omega_4$. Hence every local effect must be non-zero, {\it i.e.} $A_a\ne0$ and $B_b\ne0$. Together with the vanishing diagonal, these positive marginals allow us to apply the argument leading to Eq.~\eqref{eq:noiseless-rank-one}, which forces every local effect to have rank one.\\

If $\sum_a c_a A_a=0$ for real coefficients $c_a$, then
\begin{equation}
 \begin{aligned}
\sum_a c_a p(a,b)
&=\Tr\left[\rho_{AB}
\left(\left(\sum_a c_a A_a\right)\otimes B_b\right)\right]\\
&=0
\end{aligned}
\end{equation}
for every $b \in \Omega_4$. Equivalently, $\bm c^T P=0$, where $\bm c=(c_0,c_1,c_2,c_3)^T$. Since $P$ is invertible, $\bm c=0$.
Likewise, any relation $\sum_b d_bB_b=0$ with real coefficients $d_b$ would imply $P\bm d=0$, where $\bm d=(d_0,d_1,d_2,d_3)^T$. Invertibility of $P$ again gives $\bm d=0$. Both POVMs therefore have four
linearly independent rank-one effects and are extremal~\cite[Corollary~5]{DAriano2005}. Since a qubit PVM has at most two non-zero effects,
Corollary~\ref{cor:extremalPS} excludes projective simulability on
each side.
\end{proof}

Every genuine four-outcome extremal qubit POVM $M$ generates correlations satisfying Eq.~\eqref{eq:noiseless-four-criterion} when both parties use $M_A=M_B=M$ on a two-qubit singlet state. To see this, let $W$ be the real $4\times4$ matrix whose
$a$-th row is $\alpha_a(1,\bm n_a^T)$. The resulting probability matrix $P$ is then
\begin{equation}
  P=W\,\operatorname{diag}(1,-1,-1,-1)\,W^T.
  \label{eq:noiseless-singlet-factorisation}
\end{equation}
Since $M$ is extremal and has rank-one effects, its effects are linearly independent~\cite[Corollary~5]{DAriano2005}. Consequently, the rows of $W$ are linearly independent, because the Pauli matrices together with $\id_2$ form a basis. Thus $\rank W=4$, so $W$ and $W^T$ are invertible. Since the diagonal factor is also invertible, $P$ is invertible and therefore has rank four. The Bloch vectors of $M$ are distinct unit vectors, and $\alpha_a>0$ for every $a$,  so Eq.~\eqref{eq:prelim-noisy-target} at $v=1$ gives a probability matrix with a zero diagonal and strictly positive off-diagonal entries. In particular, using identical SIC-POVMs on a singlet give the uniform anti-correlation
\begin{equation}
  p_1^{\mathrm{tetra}}(a,b)=\frac{1-\delta_{ab}}{12}.
  \label{eq:noiseless-tetra}
\end{equation}
Thus, every genuine three- or four-outcome extremal qubit POVM on two-qubit singlet generates correlations that certify both local measurements beyond projective simulation.\\

For $\eta>0$, each matching outcome pair a positive probability, $p_v^M(a,a)=\eta\alpha_a^2>0$, so the exact-zero criteria used in the proofs no longer apply. We therefore turn to covariances constructed from the observed distribution to obtain noise-robust certification.

\section{Noise-robust certification}
\label{sec:noisy}

Depolarising noise gives the matching outcomes a non-zero probability. Certification must therefore use features of the observed distribution that persist beyond perfect anti-correlation. We show that the covariances of the outcomes provide such a test. We derive a bound on the resulting non-linear covariance witness that holds for every pair of projective-simulable qubit measurements and every shared two-qubit state.  A violation of this bound certifies that at least one local measurement lies beyond projective simulation without input randomness.

\subsection{The covariance witness}
\label{sec:covariance-witness}

To construct the covariance witness, consider the encoding where we associate each outcome $a\in\Omega_n$ of the uncharacterised POVM $M_A=\{A_a\}_{a\in\Omega_n}$  with a fixed vector
$\bm x_a\in\R^m$. These vectors  serve as numerical labels for processing the recorded outcomes. We denote the diameter of this set of vectors by
\begin{equation}
  D_{\bm x}:=\max_{a,a'}\norm{\bm x_a-\bm x_{a'}}_2.
  \label{eq:encoding-diameter}
\end{equation}

We use the same vector encoding for Bob's outcomes, so each recorded outcome pair $(a,b)$ gives a pair of vectors $(\bm x_a,\bm x_b)$. The cross-covariance matrix of the encoded outcomes under the joint distribution $P$ is

\begin{equation}
  \Gamma[P]:=\sum_{a,b}
  \bigl[p(a,b)-p_A(a)p_B(b)\bigr]\bm x_a\bm x_b^T.
  \label{eq:GammaDef}
\end{equation}

The term involving $p_A(a)p_B(b)$ subtracts the outer product of the local mean vectors, $\bigl(\sum_a p_A(a)\bm x_a\bigr)\bigl(\sum_b p_B(b)\bm x_b\bigr)^T$. For a matrix $Z$, let $s_j(Z)$ denote its $j$th singular value. We define
\begin{equation}
  \chi(Z):=\sum_j\sqrt{s_j(Z)}.
  \label{eq:chi-definition}
\end{equation}
We can compute both $\Gamma[P]$ and $\chi(\Gamma[P])$ directly from the observed probability distribution and the fixed outcome vectors. We write $\norm{Z}_{S_1}=\sum_j s_j(Z)$ for the trace norm (also called the nuclear norm) and
$\norm{Z}_{S_\infty}=\max_j s_j(Z)$ for the operator norm.\\ 

To relate the observed covariance to the local measurements, we write Alice's effects as $A_a=\alpha_a(\id_2+\bm n_a\cdot\bm\sigma)$. Weighting each effect by its outcome vector gives
\begin{equation}
  \sum_a\bm x_a\otimes A_a
=\bm c_A\otimes\id_2
+\sum_{j=1}^3(R_A)_{\cdot j}\otimes\sigma_j,
  \label{eq:response}
\end{equation}
where $\bm c_A\in\R^m$ and \emph{response matrix} $R_A\in\R^{m\times3}$. We write $(R_A)_{\cdot j}$ for the $j$th column of $R_A$.  Collecting the coefficients of the identity and the Pauli operators gives
\[
(\bm c_A)_i=\sum_a\alpha_a x_{a,i},
\qquad
(R_A)_{ij}=\sum_a\alpha_a x_{a,i}n_{a,j}.
\]
Here, $\bm x_a\otimes A_a$ denotes an $m$-component vector of qubit operators, whose $i$th component is $x_{a,i}A_a$. For a qubit state $\frac{1}{2}(\id_2+\bm u\cdot\bm\sigma)$ with Bloch vector $\bm u$, the mean encoded outcome vector of $M_A$ is $\bm c_A+R_A\bm u$. Applying the same decomposition to $\sum_b\bm x_b\otimes B_b$ defines $\bm c_B$ and $R_B$ for Bob's POVM $M_B=\{B_b\}_{b\in\Omega_n}$.

\begin{proposition}[Covariance bound for projective simulation]
\label{prop:covbound}
Fix a common outcome encoding $\{\bm x_a\}_{a\in\Omega_n}$ of diameter $D_{\bm x}$ for both parties' recorded outcomes. Every joint distribution $P$ generated by a two-qubit state $\rho_{AB}\in \mathcal{D}(\mathbb{C}^2\otimes \mathbb{C}^2)$ and local POVMs $M_A,M_B\in\PS(2,n)$ satisfies
\begin{equation}
  \chi(\Gamma[P])\le\frac{D_{\bm x}}{2}.
  \label{eq:masterbound}
\end{equation}
\end{proposition}

The bound follows by separating the contributions of the measurements
and the state. Let $R_A$ and $R_B$ be the \emph{response matrices} of Alice's and Bob's measurements, respectively, as defined in Eq.~\eqref{eq:response}. With the Bloch parameters introduced in Eq.~\eqref{eq:prelim-state-bloch}, the observed covariance factorises as
\begin{equation}
  \Gamma[P]=R_A K R_B^T,\qquad K:=T-\bm r\bm s^T.
  \label{eq:factor}
\end{equation}
We first bound the nuclear norm of the response matrix of a single qubit PVM followed by classical post-processing. Averaging the responses of the post-processed PVMs extends this bound to all projective-simulable POVMs. We next bound the operator norm of $K$ uniformly over all two-qubit states. Combining the two local response bounds with the state bound through Eq.~\eqref{eq:factor} gives Eq.~\eqref{eq:masterbound}. We give the proof in Appendix~\ref{app:covariance-proof}.\\

A violation of Eq.~\eqref{eq:masterbound} certifies that every two-qubit realisation of the distribution $P$ uses at least one local POVM outside $\PS(2,n)$. We may choose the outcome vectors using a target POVM.  Once the encoding is fixed, the bound holds for every pair of projective-simulable local qubit measurements and for every shared two-qubit state.\\

Let $P_v^M=[p_v^M(a,b)]$ be the target probability matrix in
Eq.~\eqref{eq:prelim-noisy-target} obtained by applying the same rank-one qubit POVM $M$ to both halves of a noisy singlet with visibility $v$. We encode
each outcome $a$ by its Bloch vector, $\bm x_a=\bm n_a$. For this encoding and target realisation,  Eq.~\eqref{eq:response} gives $R_A=R_B=G$ with
\begin{equation}
  G:=\sum_a\alpha_a\bm n_a\bm n_a^T,
  \qquad \Tr G=1.
  \label{eq:target-G}
\end{equation}
Here, $G$ is positive semi-definite as it is a non-negative weighted sum of outer products. Its trace is $\sum_a\alpha_a\norm{\bm n_a}_2^2$, which equals one because the Bloch vectors of the non-zero effects have unit length and the weights $\alpha_a$ sum to one by Eq.~\eqref{eq:prelim-qubit-normalisation}. For the noisy
singlet in Eq.~\eqref{eq:prelim-noisy-singlet}, $\bm r=\bm s=0$ and $K=T=-v\id_3$. Substituting into Eq.~\eqref{eq:factor} and evaluating the witness gives
\begin{equation}
  \Gamma[P_v^M]=-vG^2,
  \qquad \chi(\Gamma[P_v^M])=\sqrt v.
  \label{eq:target-covariance}
\end{equation}
Indeed, if $g_j$ are the eigenvalues of $G$, the singular values of
$-vG^2$ are the absolute values of its eigenvalues $-vg_j^2$, namely $vg_j^2$. Since $g_j\ge0$, we have $\sqrt{vg_j^2}=\sqrt v g_j$, and summing gives $\sqrt v\,\Tr G=\sqrt v$.

\subsection{Exact thresholds for trine and tetrahedral correlations}
\label{sec:symmetric-thresholds}

We first apply the witness to noisy versions of the uniform anti-correlations studied in Sec.~\ref{sec:noiseless}. For the qubit
trine POVM ($n=3$) and the qubit SIC-POVM ($n=4$), each effect of $M=\{M_a\}_{a\in \Omega_n}$ has weight $\alpha_a=1/n$, and the Bloch vectors satisfy
$\bm n_a\cdot\bm n_b=-1/(n-1)$ for $a\ne b$.
When both parties apply the same POVM $M$ to the noisy singlet of Eq.~\eqref{eq:prelim-noisy-singlet}, Eq.~\eqref{eq:prelim-noisy-target} gives the joint probabilities
\begin{equation}
  p_v^M(a,b)=
  \begin{cases}
    (1-v)/n^2, & a=b,\\[1mm]
    (n-1+v)/[n^2(n-1)], & a\ne b,
  \end{cases}
  \label{eq:noisy-symmetric-correlations}
\end{equation}
where $v=1-\eta$ is the singlet visibility.

\begin{theorem}[Exact symmetric noise thresholds]
\label{thm:symmetric-thresholds}
Let $M$ be the qubit trine or tetrahedral SIC POVM,  and let $\eta\in[0,1]$ be the white-noise fraction. The joint outcome  distribution $P_{1-\eta}^M$ defined by Eq.~\eqref{eq:noisy-symmetric-correlations} admits
a realisation with a two-qubit state and two projective-simulable
local POVMs if and only if $\eta\ge\eta_{\mathrm{crit}}(M)$, where
\begin{equation}
  \eta_{\mathrm{crit}}(M)=
  \begin{cases}
    1/4, & M\text{ is the trine POVM},\\[1mm]
    1/3, & M\text{ is the SIC  POVM}.
  \end{cases}
  \label{eq:exact-noise-thresholds}
\end{equation}
\end{theorem}

With the encoding $\bm x_a=\bm n_a$, Eq.~\eqref{eq:target-covariance} gives the witness value $\sqrt v$
for both  target correlations. The encoding diameters are $\sqrt3$ for the
trine and $\sqrt{8/3}$ for the tetrahedron. By
Eq.~\eqref{eq:masterbound}, a  two-qubit realisation with two projective-simulable
local POVMs therefore requires $v\le3/4$ for the trine and $v\le2/3$
for the tetrahedron.\\

To attain each visibility bound, Alice and Bob independently choose an unordered pair of labels  $S=\{i,j\}\in\Omega_n$ with $i\neq j$ uniformly at random. Each measures the qubit PVM along the axis defined by $\bm n_i-\bm n_j$ and reports $i$ or $j$ according to the outcome. This gives three projective branches for the trine and six for the tetrahedron. The resulting local POVM is the depolarised target measurement at visibility $t=\sqrt3/2$ for the trine and $t=\sqrt{2/3}$ for the tetrahedral SIC. The six-branch tetrahedral construction follows Ref.~\cite{Oszmaniec2017}.
Applying the resulting POVM to both halves of an ideal singlet reproduces the target correlations at visibility $v=t^2$. Further depolarising both local POVMs reproduces the target correlations at every smaller visibility.  Appendix~\ref{app:symmetric-proof} gives the constructions and
the proof of Theorem~\ref{thm:symmetric-thresholds}.

\subsection{Positive noise tolerance for extremal three- and four-outcome qubit POVMs}
\label{sec:general-robustness}

 For the trine and tetrahedron, the symmetry allows us to determine the exact noise threshold. The covariance witness also gives a strictly positive noise tolerance for each extremal qubit POVM with three or four non-zero effects. Let $M$ be such a POVM with outcome set $\Omega_n$ for $n\in\{3,4\}$. Its effects have weights
$\alpha_a>0$ and unit Bloch vectors $\bm n_a$ as in
Eq.~\eqref{eq:prelim-qubit-normalisation}. When Alice and Bob apply $M$ to a noisy singlet, their joint outcome distribution is given by Eq.~\eqref{eq:prelim-noisy-target}. Denote the diameter of the Bloch vectors of $M$ by
\begin{equation}
  D(M):=\max_{a\ne b}\norm{\bm n_a-\bm n_b}_2.
  \label{eq:DM}
\end{equation}

\begin{theorem}[Robustness of extremal qubit POVMs]
\label{thm:generalrobust}
Let $M$ be a genuine three- or four-outcome extremal qubit POVM,  and let $\eta\in[0,1]$ be the white-noise fraction of a noisy two-qubit singlet state. The witness guarantees a strictly positive noise tolerance

\begin{equation}
 \eta_{\mathrm{cert}}(M):=1-\frac{D(M)^2}{4}>0.
  \label{eq:etabound}
\end{equation}
Whenever $0\le\eta<\eta_{\mathrm{cert}}(M)$, every two-qubit realisation of the target distribution $P_{1-\eta}^M$ defined by Eq.~\eqref{eq:prelim-noisy-target} uses at least one local POVM outside $\PS(2,n)$.
\end{theorem}

With the Bloch-vector encoding $\bm x_a=\bm n_a$, Eq.~\eqref{eq:target-covariance} gives
the witness value $\chi(\Gamma[P_v^M])=\sqrt v$.  By Proposition~\ref{prop:covbound}, a two-qubit realisation with two projective-simulable local POVMs therefore requires $\sqrt v\le D(M)/2$. We use the linear independence of the effects of $M$ to show that $D(M)<2$, which gives $\eta_{\mathrm{cert}}(M)>0$.  If $\theta_{\max}$ is the largest angle between two Bloch vectors of $M$, then $D(M)=2\sin(\theta_{\max}/2)$ and the guaranteed tolerance is $\eta_{\mathrm{cert}}(M)=\cos^2(\theta_{\max}/2)$.  Appendix~\ref{app:general-proof} gives the proof.\\

For a concrete four-outcome family, take equal weights $\alpha_a=1/4$
and the four Bloch vectors obtained by choosing an even number of minus signs in
$(\pm\sqrt{q_1},\pm\sqrt{q_2},\pm\sqrt{q_3})$, where
$q_j>0$ and $q_1+q_2+q_3=1$. Appendix~\ref{app:general-proof} shows that the effects are linearly independent and that $D(M)^2=4(1-\min_j q_j)$. The resulting POVMs are therefore extremal, and their guaranteed noise tolerance is
\begin{equation}
  \eta_{\mathrm{cert}}(M)=\min\{q_1,q_2,q_3\}.
  \label{eq:non-sic-tolerance}
\end{equation}
For $(q_1,q_2,q_3)=(1/2,1/4,1/4)$, both observers apply the resulting POVM to their respective halves of a noisy singlet. Their joint outcome distribution certifies that at least one local measurement lies beyond projective simulation whenever $0\le\eta<1/4$. The family includes the tetrahedral SIC POVM at $q_1=q_2=q_3=1/3$.\\

 The single-measurement threshold $t_{\rm PM}(M)$ in Eq.~\eqref{eq:prelim-measurement-threshold} gives a sufficient condition for simulating the target correlations.  Applying $M^{(\sqrt v)}$
 to both halves of an two-qubit singlet reproduces $P_v^M$, with both local POVMs
projective-simulable whenever
\begin{equation}
  v\le t_{\rm PM}(M)^2,
  \qquad\text{or}\qquad
  \eta\ge1-t_{\rm PM}(M)^2.
  \label{eq:large-noise-simulation}
\end{equation}
As shown in Appendix~\ref{app:general-simulation}, the local response bound gives $t_{\rm PM}(M)\le D(M)/2$. For any $M$ covered by Theorem~\ref{thm:generalrobust}, let $\eta_{\mathrm{crit}}(M)$ be the smallest white-noise fraction at which $P_{1-\eta}^M$ admits a two-qubit realisation with two projective-simulable local POVMs. Theorem~\ref{thm:generalrobust} and Eq.~\eqref{eq:large-noise-simulation} give $\eta_{\mathrm{cert}}(M)\le\eta_{\mathrm{crit}}(M)\le1-t_{\rm PM}(M)^2$. For the trine and tetrahedral SIC POVMs, $t_{\rm PM}(M)=D(M)/2$, so the two bounds coincide. In the four-outcome family above, the guaranteed tolerance $\eta_{\mathrm{cert}}(M)=\min_j q_j$ approaches zero as one of the $q_j$ tends to zero.

\section{Discussion}\label{sec:discussion}
In this work, we have shown that non-projective qubit measurements can be robustly certified without relying on input randomness, using only the joint outcome distribution of local measurement devices applied to an uncharacterised shared two-qubit state. In the noiseless regime, we identified three- and four-outcome anti-correlations that certify both local measurements as lying beyond projective simulation.
To extend this to practical, noisy scenarios, we introduced a non-linear covariance witness that unifies the geometric properties of projective-simulable measurements with a universal contraction bound applicable to all two-qubit states. When both observers apply the trine POVM, or both apply the tetrahedral SIC POVM, to a noisy singlet, the exact critical white-noise fractions are $1/4$ and $1/3$, respectively. Below these thresholds, the observed correlations cannot be reproduced by any combination of a two-qubit state and projective-simulable local measurements. Furthermore, we provided explicit projective simulations that perfectly reproduce the target correlations at and above these noise boundaries. More generally, we demonstrated that our framework yields a strictly positive depolarising-noise robustness bound for correlations generated by symmetrically applying a genuine three- or four-outcome extremal qubit POVM to a noisy singlet. This bound depends on the diameter of the POVM's Bloch vectors.\\

Our findings establish an analytic framework for certifying non-projective measurement resources purely from a single joint outcome distribution. By eliminating the necessity for externally chosen preparation or measurement settings, this protocol circumvents the experimental and verification burdens associated with external seed randomness-generating devices. Previous semi-device-independent and device-independent protocols often treated input randomness as a necessary baseline resource; our results prove that the intrinsic non-classicality of generalized quantum measurements can be robustly exposed entirely through internal quantum correlations and post-processing, without external inputs. Consequently, this simplifies the measurement certification step in experiments where non-projective measurements are a resource, such as implementations of state discrimination, quantum cryptography, and certified randomness extraction where non-projective measurements provide known operational benefits.

While our covariance witness provides exact simulation thresholds for symmetric measurements like the trine and SIC POVMs, determining the precise critical noise fractions for asymmetric extremal qubit POVMs remains an open theoretical problem. For a genuine three- or
four-outcome extremal qubit POVM, our current methodology provides a positive certification bound and a complementary simulation bound which together delimit the interval in which the exact correlation threshold lies. Closing this gap to find exact thresholds for all extremal POVMs would complete the characterisation of qubit POVM robustness within this framework. 

Our framework trades the requirement of random inputs for the requirement that the local devices are not pre-correlated. This assumption rests on how the devices are built, for example, independently manufactured devices with no common classical memory. A natural extension of the present study would be to relax this independence quantitatively. The covariance witness subtracts the product of the marginals, so classically correlated mixtures of realisations can increase $\chi(\Gamma[P])$. Bounding this increase in terms of the shared randomness available, for instance, through the mutual information between the local random variables, would show how much residual correlation between the devices the certification can tolerate. Another natural direction for future research is the extension of this randomness-free certification framework beyond bipartite qubit systems to higher-dimensional qudit states or multipartite networks. Furthermore, our analysis assumes exact probability distributions, which leaves two open problems for experiments. First, $\chi(\Gamma[P])$ is nonlinear in the estimated probabilities, so a test on finite, possibly non-i.i.d., data needs a confidence bound on $\chi$. If the encoding is chosen from the same data, that bound must also account for the choice. Second, we treated only white noise. How the critical thresholds depend on partial entanglement, misalignment, and detection inefficiency remains open; for detection inefficiency this includes the best assignment of no-click events. Finally, exploring whether similar non-linear covariance witnesses can self-test specific uncharacterised states alongside the generalized measurements could further expand the operational utility of this input-free paradigm.

\section*{Acknowledgements}
We thank Michał Oszmaniec for independently raising the question of certifying measurements beyond projective simulation without input randomness while reviewing the chapter of S.R.'s thesis based on Ref.~\cite{Rout2025}. S.R. and P.H. acknowledge support from the IRA Programme, project no. FENG.02.01-IP.05-0006/23, financed by the FENG programme 2021-2027, Priority FENG.02, Measure FENG.02.01, with the support of the Foundation for Polish Science. S.S.B. is supported by the Beatriu de Pinós programme of the Ministry of Research and Universities of the Government of Catalonia (grant number 2024 BP 00267). 

\section*{LLM use disclosure}\label{sec:ai}
In the noise-robust analysis, we used ChatGPT (OpenAI) to work through the covariance factorisation and the resulting witness bound.
It also helped us check the projective simulations that attain the critical white-noise fractions of $1/4$ for the trine correlation and $1/3$ for the tetrahedral correlation. During a discussion of the noiseless trine example, ChatGPT supplied a longer derivation of extremality. The three- and four-outcome proofs presented here follows the approach we had already identified, applying the extremality criterion of D'Ariano, Lo Presti and Perinotti~\cite{DAriano2005} together with the convex-hull characterisation of projective simulability by Oszmaniec et al.~\cite{Oszmaniec2017}. We also used ChatGPT to edit the manuscript and find related work on projective simulability and covariance methods. We take full responsibility for the content of this article.

\bibliography{bib.bib}

\appendix

\section{Proofs for noise-robust certification}
\label{app:robustness}

\subsection{The covariance bound}
\label{app:covariance-proof}

Fix the outcome vectors $\bm x_a\in\R^m$ and their diameter
$D_{\bm x}$  as in  Eq.~\eqref{eq:encoding-diameter}.
We use the response matrices $R_A$ and $R_B$ defined by the decomposition in Eq.~\eqref{eq:response}.

\begin{lemma}[Local response bound]
\label{lem:response}
Every projective-simulable qubit POVM $M_A$ satisfies
\begin{equation}
  \norm{R_A}_{S_1}\le\frac{D_{\bm x}}2.
  \label{eq:nuclear-bound}
\end{equation}
\end{lemma}

\begin{proof}
By Eq.~\eqref{eq:prelim-ps-model}, the effects of a
projective-simulable qubit POVM
$M_A=\{A_a\}_{a\in\Omega_n}$ admit the decomposition
\begin{equation}
  A_a=\sum_\lambda\pi_\lambda A_a^{(\lambda)},
  \qquad
  A_a^{(\lambda)}
  :=\sum_i q(a|i,\lambda)\Pi_{i|\lambda},\label{eq:ps simulable def appendix}
\end{equation}
where $\pi_\lambda\ge0$ and $\sum_\lambda\pi_\lambda=1$.
For each $\lambda$, the effects $A_a^{(\lambda)}$ form
a POVM obtained by classically post-processing the PVM
$\{\Pi_{i|\lambda}\}_i$. For each branch $\lambda$, define $\bm c_\lambda\in\R^m$ and
$R_\lambda\in\R^{m\times3}$ by
\begin{equation}
  \begin{aligned}
    \bm c_\lambda
    &:=\frac12\sum_a
      \operatorname{Tr}(A_a^{(\lambda)})\,\bm x_a,\\
    (R_\lambda)_{:\alpha}
    &:=\frac12\sum_a
      \operatorname{Tr}(A_a^{(\lambda)}\sigma_\alpha)\,
      \bm x_a,
      \qquad \alpha=1,2,3,
  \end{aligned}
\end{equation}
where $(R_\lambda)_{:\alpha}$ denotes the $\alpha$-th
column of $R_\lambda$. The Pauli expansion of each
effect then gives
\begin{equation}
  \sum_a\bm x_a\otimes A_a^{(\lambda)}
  =
  \bm c_\lambda\otimes\id_2
  +\sum_{\alpha=1}^3
  (R_\lambda)_{:\alpha}\otimes\sigma_\alpha.
\end{equation}
Thus $R_\lambda$ is the response matrix of the branch
POVM, in the sense of Eq.~\eqref{eq:response}.
Substituting Eq.~\eqref{eq:ps simulable def appendix} into Eq.~\eqref{eq:response} and comparing the Pauli coefficients gives
\begin{equation}
  R_A=\sum_\lambda\pi_\lambda R_\lambda.
\end{equation}
If the PVM in branch $\lambda$ is trivial, its sole
non-zero projector is
$\Pi_{i_\lambda|\lambda}=\id_2$. Hence
\begin{equation}
  A_a^{(\lambda)}
  =q(a|i_\lambda,\lambda)\id_2.
\end{equation}
Since $\operatorname{Tr}(\sigma_\alpha)=0$, the definition
of $R_\lambda$ gives $R_\lambda=0$, which satisfies the
required bound. 
Otherwise, the qubit PVM has exactly two non-zero
rank-one projectors. Labelling their outcomes by
$\pm$, we write
\begin{equation}
  \Pi_{\pm|\lambda}
  =\frac{\id_2\pm\bm m_\lambda\cdot\bm\sigma}{2},
  \qquad
  \norm{\bm m_\lambda}_2=1.
\end{equation}
Conditioned on the projective outcome $\pm$, the mean
encoded output is
\begin{equation}
  \bm u_{\pm|\lambda}
  :=\sum_a q(a|\pm,\lambda)\bm x_a.
\end{equation}
Both vectors belong to $\conv\{\bm x_a\}$. Their difference is a convex combination of the vectors $\bm x_a-\bm x_{a'}$, with weights $q(a|+,\lambda)q(a'|-,\lambda)$. Each difference has norm at most $D_{\bm x}$, so the triangle inequality gives
\begin{equation}
  \norm{\bm u_{+|\lambda}-\bm u_{-|\lambda}}_2
  \le D_{\bm x}.
\end{equation}
The encoded operator of this branch is
$\bm u_{+|\lambda}\otimes\Pi_{+|\lambda}
+\bm u_{-|\lambda}\otimes\Pi_{-|\lambda}$.
Expanding the projectors and comparing the Pauli
coefficients gives
\begin{equation}
  R_\lambda
  =\frac12
  (\bm u_{+|\lambda}-\bm u_{-|\lambda})\bm m_\lambda^T.
\end{equation}
The image of $R_\lambda$ lies in the span of $\bm u_{+|\lambda}-\bm u_{-|\lambda}$, so it is at most rank one. The outer-product expression gives
\(
R_\lambda^T R_\lambda
=\frac{\norm{\bm u_{+|\lambda}-\bm u_{-|\lambda}}_2^2}{4}
\bm m_\lambda\bm m_\lambda^T
\). Since matrix $\bm m_\lambda\bm m_\lambda^T$ is a rank-one orthogonal projector, the scalar pre-factor is the largest eigenvalue of $R_\lambda^T R_\lambda$ and its square root is the largest singular value of $R_\lambda$. Since $R_\lambda$ has rank at most one, the sum of its singular values equals its largest singular value. 
Consequently,
\begin{equation}
  \norm{R_\lambda}_{S_1}
  =\frac12
  \norm{\bm u_{+|\lambda}-\bm u_{-|\lambda}}_2
  \le\frac{D_{\bm x}}2.
\end{equation}

The same bound therefore holds for every branch.  Because $\pi_\lambda\ge0$, positive homogeneity gives $\norm{\pi_\lambda R_\lambda}_{S_1}=\pi_\lambda\norm{R_\lambda}_{S_1}$. Applying the triangle inequality to $R_A=\sum_\lambda\pi_\lambda R_\lambda$ and using the bound for each branch gives

\begin{equation}
  \begin{aligned}
    \norm{R_A}_{S_1}
    &\le\sum_\lambda\pi_\lambda
      \norm{R_\lambda}_{S_1}\le\sum_\lambda\pi_\lambda\frac{D_{\bm x}}2
     =\frac{D_{\bm x}}2.
  \end{aligned}
\end{equation}
The bound holds for arbitrary branch-dependent classical post-processing.
\end{proof}

\begin{lemma}[Bound on centred state correlations]
\label{lem:statecontraction}
For every two-qubit state, $K=T-\bm r\bm s^T$ satisfies
\begin{equation}
  \norm{K}_{S_\infty}\le1.
  \label{eq:state-contraction}
\end{equation}
\end{lemma}

\begin{proof}
Let $\rho_{AB}$ be a two-qubit state with the Bloch decomposition
in Eq.~\eqref{eq:prelim-state-bloch}. We denote
$\langle O\rangle=\Tr(\rho_{AB}O)$. The reduced states have
Bloch vectors $\bm r$ and $\bm s$, with components \(r_j=\langle\sigma_j\otimes\id_2\rangle\), \(s_k=\langle\id_2\otimes\sigma_k\rangle\). The Pauli correlation matrix has entries
$T_{jk}=\langle\sigma_j\otimes\sigma_k\rangle$.
Subtracting the product of the local means gives
$K_{jk}=T_{jk}-r_js_k$.\\

For unit vectors $\bm u,\bm w\in\R^3$, consider the observables \( X=(\bm u\cdot\bm\sigma)\otimes\id_2\) and \( Y=\id_2\otimes(\bm w\cdot\bm\sigma)\). The Bloch coefficients give the means and the product moment of their outcomes
\begin{align}
  \langle X\rangle &= \bm u^T\bm r,
   \langle Y\rangle = \bm s^T\bm w, ~
  \langle XY\rangle=
  \bm u^TT\bm w.
\end{align}
The covariance of the two outcomes is therefore
\begin{equation}
  \begin{aligned}
    \operatorname{Cov}(X,Y)
    &=\langle XY\rangle-\langle X\rangle\langle Y\rangle  =\bm u^TK\bm w.
  \end{aligned}
\end{equation}

Both observables have outcomes $\pm1$, so
$\langle X^2\rangle=\langle Y^2\rangle=1$.
Applying Cauchy--Schwarz to the centred outcomes $X-\langle X\rangle$ and $Y-\langle Y\rangle$ gives
\begin{align}
  \abs{\bm u^TK\bm w}^2
  &\le \operatorname{Var}(X)\operatorname{Var}(Y)\nonumber\\
  &=\bigl[1-(\bm u^T\bm r)^2\bigr]
    \bigl[1-(\bm s^T\bm w)^2\bigr]
  \le 1.
\end{align}

For each fixed unit vector $\bm w$, Cauchy--Schwarz gives
$\abs{\bm u^TK\bm w}\le\norm{K\bm w}_2$ for every unit
vector $\bm u$. If $K\bm w\ne0$, the unit vector $\bm u=K\bm w/\norm{K\bm w}_2$ attains equality. If $K\bm w=0$, both sides vanish for every unit vector $\bm u$. Thus
\begin{equation}
  \norm{K}_{S_\infty}
  =\sup_{\norm{\bm w}_2=1}\norm{K\bm w}_2
  =\sup_{\substack{\norm{\bm u}_2=1\\\norm{\bm w}_2=1}}
    \abs{\bm u^TK\bm w}
  \le 1.
\end{equation}
\end{proof}

\begin{lemma}[Covariance factorisation]
\label{lem:factor}
Let $P$ be the joint outcome distribution generated by arbitrary
local qubit POVMs $M_A=\{A_a\}$ and $M_B=\{B_b\}$ on a two-qubit
state $\rho_{AB}$. Then $\Gamma[P]=R_A K R_B^T$. Here $R_A$ and $R_B$ are the local response matrices from Eq.~\eqref{eq:response}, and $K=T-\bm r\bm s^T$ is the centred Pauli correlation matrix of $\rho_{AB}$.
\end{lemma}

\begin{proof}
Let $p_A(a)$ and $p_B(b)$ denote the marginals of $P$. The mean encoded
outcome vectors are \(
  \bm\mu_A:=\sum_a p_A(a)\bm x_a,
  \) and \(
  \bm\mu_B:=\sum_b p_B(b)\bm x_b.
\)
For encoding coordinates $i,j$, write
\[
  F_{A,i}:=\sum_a x_{a,i}A_a,
  \qquad
  F_{B,j}:=\sum_b x_{b,j}B_b.
\]
The marginal $p_A(a)=\Tr[\rho_{AB}(A_a\otimes\id_2)]$. Using this marginal and the $i$th component of Eq.~\eqref{eq:response}, we obtain
\begin{align*}
  (\bm\mu_A)_i
  &=\Tr[\rho_{AB}(F_{A,i}\otimes\id_2)]\\
  &=(\bm c_A)_i+\sum_\alpha(R_A)_{i\alpha}r_\alpha
   =(\bm c_A+R_A\bm r)_i.
\end{align*}
Thus $\bm\mu_A=\bm c_A+R_A\bm r$, and similarly
$\bm\mu_B=\bm c_B+R_B\bm s$. By Eq.~\eqref{eq:GammaDef}, $\Gamma[P]=C-\bm\mu_A\bm\mu_B^T$,
where $C:=\sum_{a,b}p(a,b)\bm x_a\bm x_b^T$ is the
uncentred second moment matrix. Its $(i,j)$ entry is
\begin{align}
  C_{ij}
  &=\sum_{a,b}p(a,b)x_{a,i}x_{b,j} =\Tr[\rho_{AB}(F_{A,i}\otimes F_{B,j})]
\end{align}
where the second equality follows from the Born rule and
bilinearity of the tensor product. Expanding both operators
using Eq.~\eqref{eq:response} and evaluating the Pauli moments of $\rho_{AB}$ gives
\begin{align*}
  C_{ij}
  ={}&(\bm c_A)_i(\bm c_B)_j
      +(\bm c_A)_i(R_B\bm s)_j\\
     &+(R_A\bm r)_i(\bm c_B)_j
      +\sum_{\alpha,\beta}
       (R_A)_{i\alpha}T_{\alpha\beta}(R_B)_{j\beta}.
\end{align*}
The last sum is $(R_ATR_B^T)_{ij}$. Thus
\begin{align}
  C
  ={}&\bm c_A\bm c_B^T+\bm c_A(R_B\bm s)^T +(R_A\bm r)\bm c_B^T+R_A T R_B^T.
\end{align}
Subtracting $\bm\mu_A\bm\mu_B^T=(\bm c_A+R_A\bm r)(\bm c_B+R_B\bm s)^T$ cancels the three terms involving $\bm c_A$ or $\bm c_B$. The remaining terms give
\begin{align*}
  \Gamma[P]
  &=R_A T R_B^T-(R_A\bm r)(R_B\bm s)^T =R_A(T-\bm r\bm s^T)R_B^T\\
  &=R_A K R_B^T.
\end{align*}
\end{proof}

\begin{proof}[Proof of Proposition~\ref{prop:covbound}]
By definition, the Schatten-$1/2$ quasi-norm satisfies
$\norm{Z}_{S_{1/2}}=(\sum_j\sqrt{s_j(Z)})^2=\chi(Z)^2$.
We use the Schatten H\"older inequality
$\norm{XY}_{S_p}\le\norm{X}_{S_q}\norm{Y}_{S_r}$
for $1/p=1/q+1/r$ with positive finite exponents~\cite[Exercise~IV.2.7, Eq.~(IV.43)]{Bhatia1997}. For compatible rectangular matrices, the inequality follows by padding both factors with zeros to a common square size.
The case $(p,q,r)=(1/2,1,1)$ bounds the product of
$R_AK$ and $R_B^T$. The endpoint  $(p,q,r)=(1,1,\infty)$
bounds $\norm{R_AK}_{S_1}$, and follows by taking $r\to\infty$ with $q=1$.
Using Lemma~\ref{lem:factor} and the invariance of the
nuclear norm under transposition, we obtain
\begin{align*}
  \chi(\Gamma[P])^2
  &=\norm{R_A K R_B^T}_{S_{1/2}}\le\norm{R_A K}_{S_1}\norm{R_B^T}_{S_1}\\
  &\le\norm{R_A}_{S_1}\norm{K}_{S_\infty}\norm{R_B}_{S_1}
  \le\frac{D_{\bm x}^2}{4}.
\end{align*}
The last inequality uses the local response bounds from
Lemma~\ref{lem:response} and the state bound from
Lemma~\ref{lem:statecontraction}. Taking square roots
gives Eq.~\eqref{eq:masterbound}.
\end{proof}

\subsection{Exact thresholds for trine and tetrahedral correlations}
\label{app:symmetric-proof}

\begin{proof}[Proof of Theorem~\ref{thm:symmetric-thresholds}]
Let $n=3$ for the trine POVM and $n=4$ for the tetrahedral SIC POVM.
In both cases, the effects have weights $\alpha_a=1/n$, and the unit Bloch vectors satisfy
$\bm n_a\cdot\bm n_b=-1/(n-1)$ for $a\ne b$. For each POVM, the distance between any two distinct Bloch vectors is
\begin{equation}
  D_n=\sqrt{\frac{2n}{n-1}}.
\end{equation}
By Eq.~\eqref{eq:target-covariance}, the encoding $\bm x_a=\bm n_a$
gives the witness value $\chi(\Gamma[P_v^M])=\sqrt v$. For any realisation of $P_v^M$ with a two-qubit state and two projective-simulable local POVMs, Proposition~\ref{prop:covbound} therefore requires $\sqrt v\le D_n/2$. Squaring gives
\begin{equation}
  v\le\frac{D_n^2}{4}=\frac{n}{2(n-1)}.
  \label{eq:symmetric-necessary-visibility}
\end{equation}
This gives $v\le3/4$ for the trine and $v\le2/3$ for the tetrahedral
correlation.\\

To attain the visibility bound in Eq.~\eqref{eq:symmetric-necessary-visibility},
each observer independently chooses a two-element subset $S=\{i,j\}\subseteq\Omega_n$
uniformly from the $\binom n2$ possibilities and measures the PVM
\begin{equation}
  \Pi_i^{(S)}=\frac12\left[
    \id_2+\frac{\bm n_i-\bm n_j}{D_n}\cdot\bm\sigma
  \right],
  \qquad
  \Pi_j^{(S)}=\id_2-\Pi_i^{(S)}.
  \label{eq:symmetric-projective-branches}
\end{equation}
The observer reports  $i$ or $j$ according to whether the outcome corresponds to $\Pi_i^{(S)}$ or $\Pi_j^{(S)}$. Only pairs
containing $i$ contribute to the effective POVM effect $F_i$, and
each such pair is chosen with probability $1/\binom n2$. Completeness and
equal weights give $\sum_j\bm n_j=0$, so
$\sum_{j\ne i}(\bm n_i-\bm n_j)=n\bm n_i$. Hence
\begin{align}
  F_i
  &=\frac{1}{\binom n2}\sum_{j\ne i}\Pi_i^{(\{i,j\})}\nonumber\\
  &=\frac{1}{n(n-1)}\sum_{j\ne i}
    \left[\id_2+
    \frac{\bm n_i-\bm n_j}{D_n}\cdot\bm\sigma\right]\nonumber\\
  &=\frac1n\left[\id_2+
    \frac{n}{(n-1)D_n}\bm n_i\cdot\bm\sigma\right]
   =M_i^{(t_n)},
\end{align}
where the measurement visibility is
\begin{equation}
  t_n:=\frac{n}{(n-1)D_n}
      =\frac{D_n}{2}
      =\sqrt{\frac{n}{2(n-1)}}.
  \label{eq:symmetric-measurement-visibility}
\end{equation}
For $n=3$, this construction uses a uniform mixture of three PVMs. For $n=4$, it gives the six-branch tetrahedral simulation of
Ref.~\cite{Oszmaniec2017}. For any $0\le t\le t_n$, the depolarised measurement has effects
\begin{equation}
  M_i^{(t)}=\frac{t}{t_n}F_i+
     \left(1-\frac{t}{t_n}\right)\frac{\id_2}{n}.
\end{equation}
Each observer can implement this projective-simulable measurement
by using the PVM mixture above with probability $t/t_n$ and otherwise
reporting a uniformly sampled label. If $v\le t_n^2=\frac{D_n^2}{4}$, both observers
choose $t=\sqrt v$ and apply $M^{(t)}$ to their respective halves of
the ideal two-qubit singlet. Writing $a$ and $b$ for Alice's and Bob's reported
outcomes, Eq.~\eqref{eq:prelim-visibility-product} gives
\begin{equation}
  \Tr\!\left[\proj{\psi^-}
       \bigl(M_a^{(t)}\otimes M_b^{(t)}\bigr)\right]
  =p_{t^2}^M(a,b)=p_v^M(a,b).
\end{equation}
All random choices are local and independent between observers.
The construction therefore reproduces $P_{1-\eta}^M$ with two
projective-simulable local POVMs for every
$1-t_n^2\le\eta\le1$, including the threshold. Together with
Eq.~\eqref{eq:symmetric-necessary-visibility}, this proves that the
exact white-noise threshold is $\eta_{\mathrm{crit}}(M)=1-t_n^2=1-\frac{D_n^2}{4}$,
equal to $1/4$ for the trine and $1/3$ for the tetrahedral SIC POVM.
\end{proof}

\subsection{General extremal POVMs and positive tolerance}
\label{app:general-proof}

\begin{proof}[Proof of Theorem~\ref{thm:generalrobust}]
Let $M$ be an extremal qubit POVM with $n\in\{3,4\}$ non-zero effects. Its non-zero effects are rank one and are linearly independent~\cite[Corollary~5 and Sec.~3]{DAriano2005}, so
\begin{equation}
  M_a=\alpha_a(\id_2+\bm n_a\cdot\bm\sigma),
  \qquad \alpha_a>0,\quad \norm{\bm n_a}_2=1,
\end{equation}
with $\sum_a\alpha_a=1$ and $\sum_a\alpha_a\bm n_a=0$.
Equation~\eqref{eq:target-covariance} gives $\chi(\Gamma[P_v^M])=\sqrt v$
for the encoding vector $\bm x_a=\bm n_a$. The encoding diameter is $D(M)$. Proposition~\ref{prop:covbound} consequently requires
$v\le D(M)^2/4$ for every two-qubit realisation in which both local POVMs belong to $\PS(2,n)$. Since $v=1-\eta$,   $0\le\eta<1-D(M)^2/4$ excludes every such realisation.

To prove strict positivity, note that $D(M)\le2$ because the Bloch vectors have unit length. Suppose equality holds. Two unit Bloch vectors
would then be antipodal, say $\bm n_j=-\bm n_i$, implying
\begin{equation}
  \id_2=\frac{M_i}{2\alpha_i}+\frac{M_j}{2\alpha_j}.
\end{equation}
Subtracting this identity from $\sum_aM_a=\id_2$ gives the relation
\begin{equation}
0=\sum_{a\notin\{i,j\}}M_a
+\left(1-\frac{1}{2\alpha_i}\right)M_i
+\left(1-\frac{1}{2\alpha_j}\right)M_j.
\end{equation}
Since $n\ge3$, at least one effect occurs in the first sum with coefficient $1$. The relation is therefore non-trivial, contradicting linear independence. Thus $D(M)<2$, which implies $\eta_{\mathrm{cert}}(M)>0$.
\end{proof}

For the four-outcome family in Sec.~\ref{sec:general-robustness}, the four
Bloch vectors are the rows of
\begin{equation}
  N=\begin{pmatrix}
    \sqrt{q_1}& \sqrt{q_2}& \sqrt{q_3}\\
    \sqrt{q_1}&-\sqrt{q_2}&-\sqrt{q_3}\\
   -\sqrt{q_1}& \sqrt{q_2}&-\sqrt{q_3}\\
   -\sqrt{q_1}&-\sqrt{q_2}& \sqrt{q_3}
  \end{pmatrix}.
  \label{eq:non-sic-vectors}
\end{equation}
Each row has unit norm because $\sum_j q_j=1$, and the rows sum to zero because each column has two positive and two negative entries of equal magnitude. With equal weights $\alpha_a=1/4$, these vectors therefore define a rank-one qubit POVM. Since $N^TN=4\operatorname{diag}(q_1,q_2,q_3)$ and $q_j>0$, the columns of $N$ are linearly independent. The zero row sum makes the all-ones column $\bm 1$ orthogonal to each of them, so the matrix $[\bm 1|N]$ is invertible. The coefficient rows of the effects in the Pauli basis are $\tfrac14(1,\bm n_a^T)$, so the effects are linearly independent. Since they are rank one, the POVM is extremal.\\

Each pair of distinct rows agrees in one coordinate and differs in the other two. If the rows agree in coordinate $j$, their inner product is $q_j-\sum_{k\ne j}q_k=2q_j-1$, so their squared distance is $4(1-q_j)$. Each coordinate occurs as the coordinate of agreement for some pair. Taking the largest squared distance gives
\begin{equation}
  D(M)^2=4\bigl(1-\min_jq_j\bigr),
  \qquad \eta_{\mathrm{cert}}(M)=\min_jq_j.
\end{equation}
All pairwise inner products equal $-1/3$ precisely when
$q_1=q_2=q_3=1/3$, giving the tetrahedral SIC POVM.\\

\subsection{Projective simulations at larger noise}
\label{app:general-simulation}

We justify Eq.~\eqref{eq:large-noise-simulation} for an arbitrary
genuine three- or four-outcome extremal qubit POVM $M$.
When $v\le t_{\rm PM}(M)^2$, the measurement
$M^{(\sqrt v)}=\Phi_{\sqrt v}(M)$ belongs to $\PS(2,n)$ because $M^{(t)}$ depends affinely on $t$, the POVMs at $t=0$ and $t=t_{\rm PM}(M)$ are projective-simulable, and $\PS(2,n)$ is convex. Applying this measurement to both halves of an ideal singlet gives
\begin{align}
  \Tr\!\left[\proj{\psi^-}
    (M_a^{(\sqrt v)}\otimes M_b^{(\sqrt v)})\right]
  &=\alpha_a\alpha_b
    (1-v\bm n_a\cdot\bm n_b)\nonumber\\
  &=p_v^M(a,b).
\end{align}
Thus $\eta\ge1-t_{\rm PM}(M)^2$ suffices for a realisation with two projective-simulable local measurements.\\

More generally, suppose a two-qubit realisation with two projective-simulable local POVMs  reproduces $P_v^M$. For $\lambda\in[0,1]$, Alice uses private randomness to replace her output with an independent sample from $(\alpha_a)_{a\in\Omega_n}$ with probability $1-\lambda$. Bob's marginal in $P_v^M$ is $(\alpha_b)_{b\in\Omega_n}$, so the replacement branch gives the product distribution $P_0^M=[\alpha_a\alpha_b]$. The resulting distribution is $\lambda P_v^M+(1-\lambda)P_0^M=P_{\lambda v}^M.$ This local post-processing preserves projective simulability. Hence every smaller
visibility is also simulable. The set of simulable distributions is compact as the image of the compact product $\mathcal D(\mathbb C^2\otimes\mathbb C^2)\times\PS(2,n)^2$ under the continuous Born-rule map.  Since $P_{1-\eta}^M$ depends continuously on $\eta$ and $P_0^M$ is simulable, the noise fractions admitting a simulation form a non-empty closed interval $[\eta_{\mathrm{crit}}(M),1]$.\\

For the encoding $\bm x_a=\bm n_a$, the response matrix of
$M^{(t)}$ is $tG$, since depolarisation multiplies each Bloch vector by $t$. Since $G\ge0$ and $\Tr G=1$,
$\norm{tG}_{S_1}=t$. Applying Lemma~\ref{lem:response} at $t=t_{\rm PM}(M)$ gives
\begin{equation}
  t_{\rm PM}(M)\le\frac{D(M)}2.
\end{equation}
For the trine and tetrahedral SIC POVMs, the construction at visibility $t_n$ in Eq.~\eqref{eq:symmetric-measurement-visibility} attains this local bound. Applying the resulting POVM to both halves of the singlet gives correlation visibility $t_n^2$, which attains the bound in Eq.~\eqref{eq:symmetric-necessary-visibility}. Thus, for these two POVMs,
\begin{equation}
  t_{\rm PM}(M)=\frac{D(M)}2,
  \qquad \eta_{\mathrm{crit}}(M)=1-t_{\rm PM}(M)^2.
\end{equation}

\end{document}